\documentclass[twocolumn,aps,prl,showpacs,amsmath,amssymb,floatfix]{revtex4-1}
\usepackage{graphicx}
\usepackage{dcolumn}
\usepackage{bm,xcolor}

\begin{document}

\title{Emergent Bistability and Switching in a Nonequilibrium Crystal}

\author{Guram Gogia}
\author{Justin C. Burton}
\affiliation{Department of Physics, Emory University, Atlanta, GA, 30322}
\date{\today}
 
\begin{abstract}
Multistability is an inseparable feature of many physical, chemical and biological systems which are driven far from equilibrium. In these nonequilibrium systems, stochastic dynamics often induces switching between distinct states on emergent timescales, for example, bistable switching is a natural feature of noisy, spatially-extended systems that consist of bistable elements. Nevertheless, here we present experimental evidence that bistable elements are not required for the global bistability of a system. We observe temporal switching between a crystalline, condensed state and a gas-like, excited state in a spatially-extended, quasi-two-dimensional system of charged microparticles. Accompanying numerical simulations show that conservative forces, damping, and stochastic noise are sufficient to prevent steady-state equilibrium, leading to switching between the two states over a range of time scales, from seconds to hours. 
\end{abstract}
\maketitle

Systems driven far from equilibrium, either through external fields or from active constituents, exhibit a diverse range of collective phenomena. Although no universal framework exists for describing nonequilibrium systems, in many cases a steady state is reached where some aspects of equilibrium theories can be applied \cite{Mukamel2000}. However, nonequilibrium systems often never reach a steady state, and constantly switch behavior \cite{Hinrichsen2000}. In biological systems, switching is a common theme in the regulation of the cell cycle \cite{Pomerening2003}, gene expression \cite{Maamar2007}, and neural activity \cite{Pigorini2015}. Certain system features can affect switching behavior over time, for example, stochastic noise can both facilitate and hinder switching \cite{Lindner2004}. The role of noise is often discussed in terms of stochastic resonance or stochastic facilitation, where noise serves a constructive role to amplify switching between two states \cite{Gammaitoni1998,McNamara1988,McDonnell2011}. 

For spatially-extended systems such as a neural network, the existence of collective switching between states is supported by the bistable nature of the individual neurons. Other well-known examples include arrays of diode resonators \cite{Locher1996} and nematic liquid crystals \cite{Sharpe2003}. More interestingly, bistable switching can emerge from collective system behavior, even if the system constituents are not bistable. Bistability has been demonstrated in several numerical studies on collective systems composed of monostable elements \cite{Vilar1997,Cubero2009,Miloshevich2009}. However, observations of similar behavior in experimental systems are lacking. Using an electrostatically-levitated, quasi-two-dimensional collection of charged microparticles, here we show how dynamical switching between distinct states can emerge as a collective system behavior rather than a consequence of bistable elements. 

\begin{figure*}
\includegraphics[width=6.6 in]{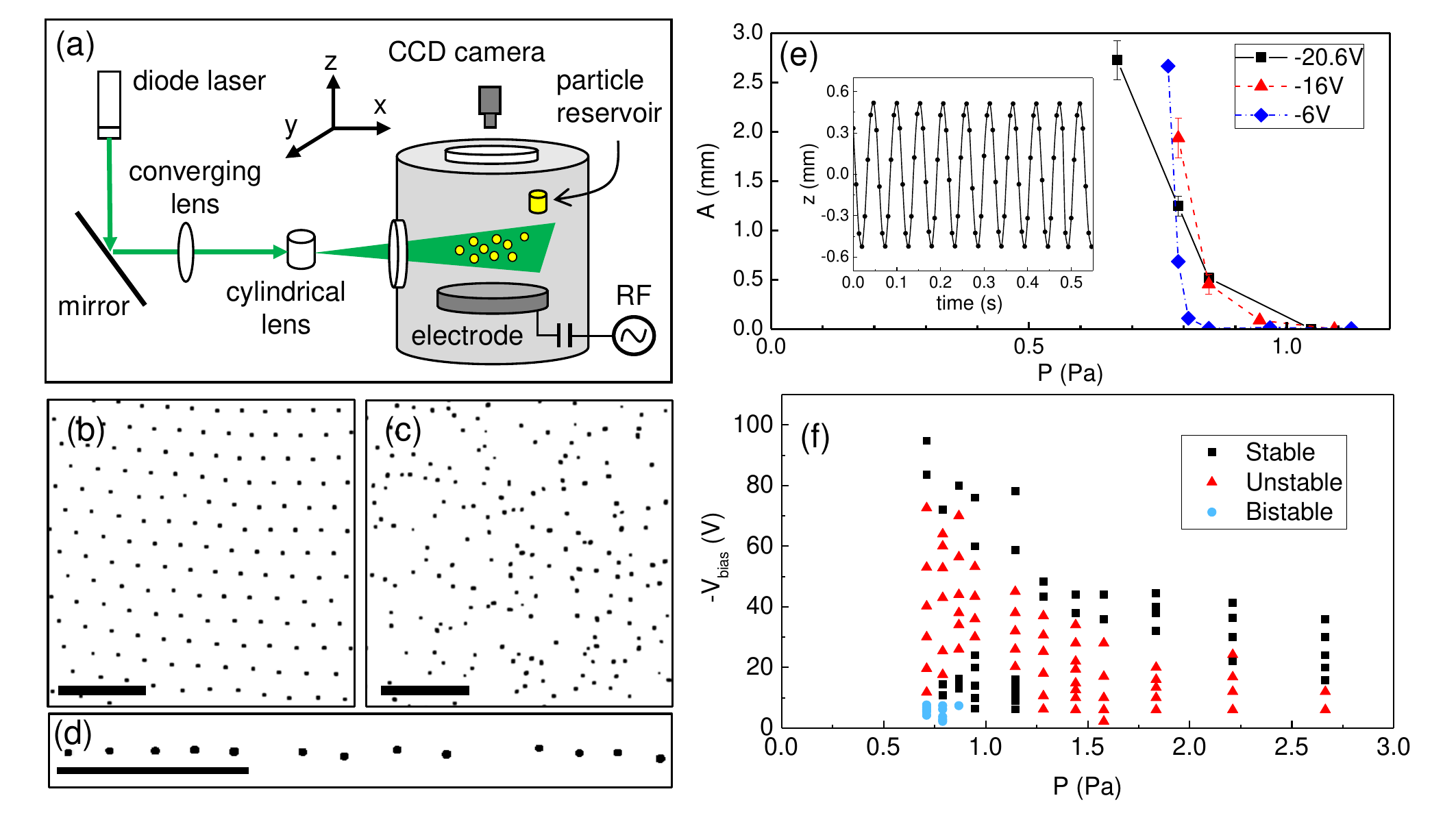}
\caption[]{(a) Diagram of the experimental setup. The particles were introduced into the vacuum chamber by mechanically shaking the particle reservoir. The particles were illuminated using vertical and horizontal (not shown) laser sheets and imaged using two cameras, one from above and one from the side. Central regions of crystalline (b) and gas-like (c) samples. (d) Side view of the particle monolayer. Scale bars correspond to 5 mm. (e) Amplitude of vertical oscillation for a single particle as a function of $P$ for different values of $V_{bias}$. \textit{Inset}: Evolution of the $z$-position of a single particle. (f) Phase diagram of a system consisting of 691 melamine-formaldehyde particles.  
\label{setup}} 
\end{figure*}

\textit{Experiments} --- Figure \ref{setup}a shows the experimental conditions for particle levitation. The micron-scale particles are levitated in a weakly-ionized argon plasma \cite{Shukla2009,Merlino2004}. Direct visualization of the individual particles allows using such ``dusty'' or complex plasmas as model systems for studying various condensed matter phenomena \cite{Joyce2002,Sutterlin2009}. Gas pressure and plasma density control the residual charge on the particles, and thus their ability to form a crystalline lattice with little or no Brownian motion (Fig.\ \ref{setup}b), or a kinetic gas-like state (Fig.\ \ref{setup}c). The particles interact via a repulsive, screened Coulomb potential, where the typical screening length is $\lambda_D\approx$ 1 mm. The particles levitate at a vertical position defined by a balance between gravity and electrostatic repulsion from the negatively-biased aluminum electrode at voltage $V_{bias}$ (Fig.\ \ref{setup}d). Weak horizontal confinement is provided by a gradient in ion density which keeps the negatively-charged particles near the center of the electrode. 

Our experiments use melamine-formaldehyde particles obtained from microParticles GmbH. The mean diameter of the particles was 9.46 $\mu$m, with a coefficient of variation, $c_v \approx 1.1\%$. This inherent polydispersity led to a variation in the equilibrium height of the particles and their corresponding vertical oscillation frequency. To remove ``dimer" particles consisting of two particles stuck together, we pulsated the plasma using the duty cycle on the RF power supply (Seren model R301, 13.56 MHz). The duty cycle was adjusted so that heavier particles impacted the electrode, while smaller particles were levitated once the plasma was re-initiated. The bias voltage on the electrode was adjusted by varying the RF power ($\sim$ 2-10 W) delivered to the plasma. Lower voltages were achieved by detuning the matching network to reduce the power. A 300 mW diode laser was used as a source for the horizontal and vertical laser sheets (Laserglow Technologies). The thickness of the laser sheet was $\approx$ 100 $\mu$m. The particle motion was imaged from above and from the side using USB 3.0 digital video cameras (Point Grey). We tracked the particle positions using a python-based software package \cite{Allan2016}. Particle velocities were calculated by subsequent differences in position in each frame.  

\begin{figure*}
	\begin{center}
		\includegraphics[width=6.3 in]{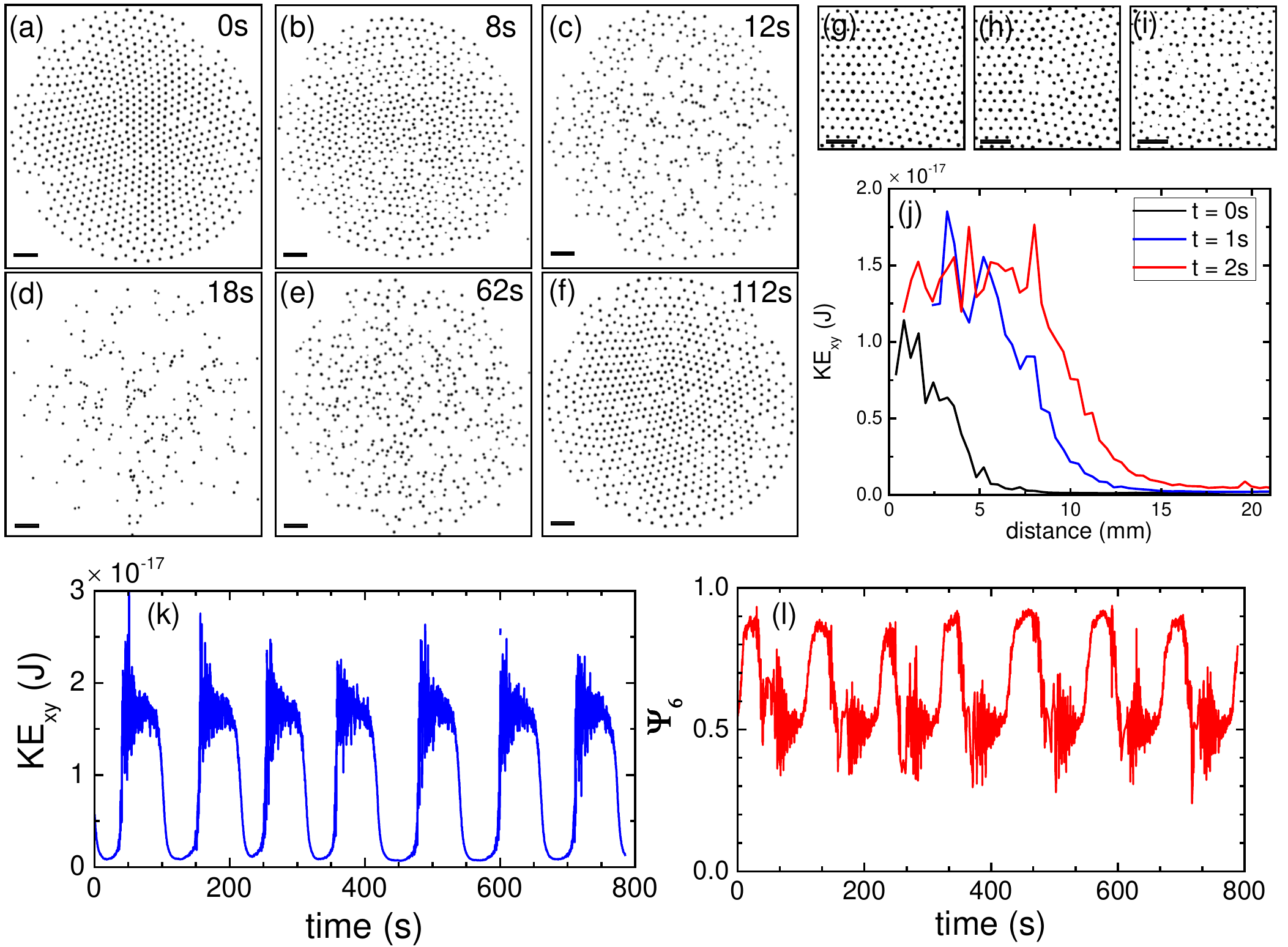}
		\caption[]{(a)-(f) Image sequence showing one cycle of melting and recrystallization of a particle monolayer made from 691 particles. Some particles move out of the plane of the laser sheet during the vertical oscillations and instability periods so they are not visible. (g)-(i) Sequence of images showing the nucleation of melting taking place in the central part of the sample. The time between frames is 1 s. (j) Average horizontal kinetic energy per particle as a function of distance from the nucleation event. (k) Average horizontal kinetic energy per particle and (l) six-fold bond orientational order parameter for $P$ = 0.71 Pa and $V_{bias}$ = -6 V. The scale bars in all images correspond to 5 mm. The large noise visible during instability periods is due to difficulties in tracking highly-energetic particles.
			\label{ExpResults} }
	\end{center} 
\end{figure*}

Often condensed matter systems have some intrinsic disorder and fluctuations. In complex plasmas, variations in particle size introduce quenched disorder, leading to non-reciprocal interactions \cite{Ivlev2015} and melting of the crystal under weak confinement \cite{Couedel2010,Liu2010}. More importantly for our experiments, single particles experience self-induced vertical oscillations at low gas pressures. A number of proposed mechanisms for the source of the vertical oscillations have been suggested, initially, a delayed charging of the particles \cite{Nunomura1999} and subsequently, spatial charge variation \cite{Samarian2001} and plasma sheath inhomogeneities \cite{Resendes2002}. Although there is no consensus on the source of the oscillations, they are crucial to the emergence of bistability in our experiments. Figure \ref{setup}e shows the amplitude of vertical oscillation for a single particle as a function of pressure, $P$, for two different values of $V_{bias}$. The inset shows the $z$-position of a particle in time, where the characteristic oscillation frequency is $\approx$ 20 Hz. 

By varying $P$ and $V_{bias}$, which control the dissipative and interaction forces, respectively, we explore the system behavior (Fig.\ \ref{setup}f). A similar phase diagram for smaller particles is shown in Fig.\ S1. For large values of $P$ and $-V_{bias}$, the particles form a stable crystalline monolayer. Initially, reducing $P$ or $-V_{bias}$ destabilizes the monolayer as small fluctuations become more pronounced. However, at low $P$ and $-V_{bias}$, $\lambda_D$ increases due to the low plasma density, leading to an increase in the inter-particle distance and the formation of stable crystals. Surprisingly, in this regime we also observe a ``bistable'' phase, where the particle monolayer continually switches between a crystalline state and a gas-like state on a minutes-long timescale. This timescale is not inherent in the dynamics of individual particles, but rather emerges from the many-body dynamics of a stochastically-driven system with damping.

Figures \ref{ExpResults}a-f show images from a typical bistable sample over 112 s (see Supplementary Videos 1 and 2 for the full evolution of similar samples). The vertical oscillations play a key role here. As the particles begin to oscillate vertically, they move in-and-out of the laser sheet, as evidenced by the apparent ``holes'' in the crystalline sample. Most of the particles oscillate in a synchronous fashion (see Supplementary Video 3 for a side view). Since each particle's oscillation frequency is slightly different due to variations in particle mass, some particles begin to move out-of-phase with the rest of the sample. As the amplitude of vertical oscillation increases, nonlinear interactions can lead to a nucleation of melting in the sample, where energy is rapidly redistributed into the horizontal degrees of freedom. Figures \ref{ExpResults}g-i show a typical nucleation event in the sample. The redistribution of kinetic energy spreads at the speed of sound in the crystalline lattice, roughly 1 cm/s (Fig.\ \ref{ExpResults}j), and is reminiscent of similar phenomena observed in vibrated granular layers \cite{Losert1999}.

After a nucleation event, the particles form a highly-energetic, gas-like state. However, hydrodynamic damping causes the kinetic energy to decay and consequently the particles are able to re-crystallize until the vertical oscillation of particles begins again, and the cycle repeats itself for hours without modifying any external parameters (see Fig.\ S2). Switching between the two states occurs over a wide range of time scales, depending on the environmental conditions and sample size, although it is often quasi-periodic. Figure \ref{ExpResults}k shows the horizontal kinetic energy per particle, $KE_{xy}$ over a few bistable periods. 
This switching can also be observed in the temporal evolution of structural order of the sample, as evidenced by the six-fold bond orientational order parameter \cite{Brock1986}, $\Psi_6$, which is equal to $1$ for a perfect hexagonal crystal and 0 for a completely disordered sample (Fig.\ \ref{ExpResults}l). 

\textit{Numerical Simulations} --- In order to understand the essential ingredients for the emergence of bistability in our system, we constructed a numerical simulation based on our experiments. The simulation consists of interacting charged particles with vertical and horizontal electrostatic confinement, resulting in a system of conservative forces. We add two other forces on each particle: a hydrodynamic drag force as given by the Epstein law \cite{Shukla2009}, which relates the damping experienced by a particle to the ambient pressure, and a spatially-uniform, Langevin force in the vertical direction with relative strength $\beta$. The amplitude of oscillation is determined by a balance between the stochastic forcing, $\beta$, and dissipation due to damping, $\gamma$. As discussed above, the nature of the vertical oscillations is not well-understood.  We initially chose to drive the system using periodic forcing in the vertical direction, however, this did not cause bistable switching. This is likely due to the fact that a sinusoidally-driven harmonic oscillator responds at the driving frequency, so that all particles move together, whereas a stochastically-driven harmonic oscillator has a peak response at its fundamental frequency, leading to out-of-phase oscillations between neighboring particles and melting of the crystal (Fig.\ \ref{ExpResults}g-i).

The numerical simulations used a custom molecular dynamics code which treated particles as point-like spherical capacitors where the charge is proportional to the particle voltage, $V_p$. Particle diameters were chosen from a Gaussian distribution with mean = 10 $\mu$m and coefficient of variation = $c_v$. The quenched disorder in the system was due to variation in both mass and charge among the particles. The inter-particle potential was $U(r_{ij})=q_{i}q_{j}e^{-r_{ij}/\lambda_D}/(4\pi\epsilon_0 r_{ij})$, where $q$ is the charge on each particle, and $r_{ij}$ is the distance between the centers of particle $i$ and $j$, and $\lambda_D= 1$ mm. The vertical confining potential was $V_v=-\chi_v z^2/2$ and the horizontal confining potential was $V_h=-\chi_h (x^2+y^2)/2$, where $\chi_v = 4.13\times10^6$ V/m$^2$ and $\chi_h=2000$ V/m$^2$. These values were chosen to satisfy the typical oscillation frequencies of particles in the experiment (i.e. 20 Hz for vertical oscillations and 0.5 Hz for horizontal oscillations). The hydrodynamic drag force was $\vec{\bf{F}}_d=-\gamma m \vec{\bf v}$, where $m$ is the particle mass and $\gamma$ is the damping coefficient. The stochastic force in the vertical direction used to model the self-induced vertical oscillations was given by $\vec{\bf F}_s=\eta(\beta,t) m \sqrt{\Delta t_0/\Delta t}$ $\hat{\bf z}$, where $\eta(\beta,t)$ is a random number chosen at each time step from a Gaussian distribution with standard deviation $\beta$, $\Delta t$ is the time step in the simulation, and $\Delta t_0=0.001$ s. The particle positions and velocities were advanced in time using velocity-Verlet integration. Given the large parameter space, we used $\gamma$ = 0.2 s$^{-1}$, $\beta$ = 0.5 m/s$^2$, $V_p$ = -6 V, and $c_v$ = 1.25\% as default parameters unless otherwise noted.

\begin{figure}[!t]
	\begin{center}
		\includegraphics[width=3.3 in]{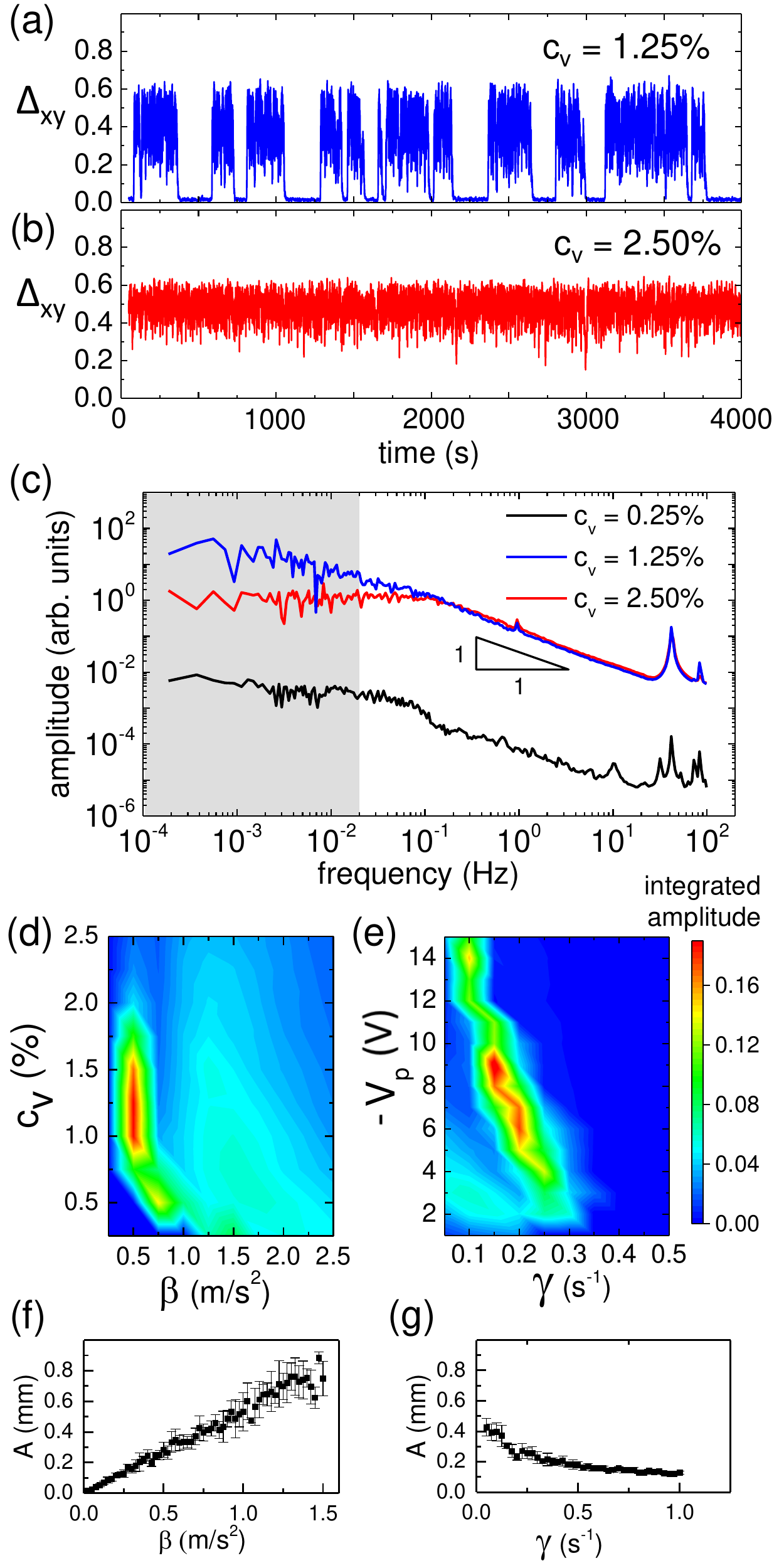}
		\caption[]{(a-b) Temporal evolution of $\Delta_{xy}$ for two separate values of $c_v$. (c) Fourier transform of $\Delta_{xy}$ from (a-b), and one for $c_v$ = 0.25\%. (d) Color map showing the integrated amplitude of the Fourier spectrum below 0.02 Hz, as shown by the gray region in (c), as a function of $\beta$ and $c_v$. (e) Color map showing the integrated Fourier amplitude as a function of $\gamma$ and $V_p$. (f-g) Average vertical oscillation amplitude of a single particle as a function of $\beta$ and $\gamma$.
			\label{SimulResults}} 
	\end{center} 
\end{figure}

There are two system properties which facilitate bistable switching. The first is quenched disorder due to variation of particles sizes, given by $c_v$. The second is dynamical noise derived from the vertical stochastic forcing, given by $\beta$. Figure \ref{SimulResults}a-b shows the time evolution of the ratio of horizontal to total kinetic energy, $\Delta_{xy}=KE_{xy}/(KE_z+KE_{xy})$, for two values of $c_v$. For $c_v = 1.25\%$, which is close to the experimental value, we observe bistable switching. For larger values of $c_v$, the system remains unstable and the energy is nearly equipartitioned between the various degrees of freedom, which corresponds to $\Delta_{xy}=2/3$. For $c_v\lesssim 0.75\%$, the particles oscillate in-phase and the sample remains crystalline. The switching timescales can be extracted from the Fourier spectrum of $\Delta_{xy}$, as shown in Fig.\ \ref{SimulResults}c. Bistable switching corresponds to a large Fourier amplitude at low frequencies. If the system was self-organized near a critical point in the dynamics, then one would expect a 1/$f$ behavior in the Fourier spectrum \cite{Bak1987,Kadanoff1989,Manna1991}. The data shows good agreement with this scaling at higher frequencies, but rolls off at low frequencies.

By varying both $\beta$ and $c_v$, we find that the intensity of switching is maximized for intermediate values of both quenched disorder and dynamical noise. Figure \ref{SimulResults}d shows a heat map of the integrated Fourier amplitude below $f=0.02$ Hz, which we use as a proxy for the intensity of switching. The map verifies that too little noise is insufficient to induce switching between the states, whereas too much noise leaves the system stuck in the gas-like state. However, $\beta$ and $c_v$ can not be easily controlled in the experiment. In order to better connect the simulation with the experiment, in Fig.\ \ref{SimulResults}e we plot the switching intensity as a function of $V_{P}$ and $\gamma$. The switching occurs for a fairly narrow range of $V_{P}$ and $\gamma$, which is consistent with Fig.\ \ref{setup}f. The agreement is good considering that in the experiment, changing $V_{bias}$ also changes the vertical and horizontal confinement. Finally, in Figs.\ \ref{SimulResults}f-g, we plot the average vertical oscillation amplitude of a single particle as a function of $\beta$ and $\gamma$, which also shows good agreement with the typical amplitudes measured in the experiment (Fig.\ \ref{setup}e).

\textit{Summary} --- The experiments and simulations presented here display a broad class of nonequilibrium phenomena in a single system with minimal ingredients and rich dynamics. We have experimentally demonstrated global bistability in a spatially-extended system composed of non-bistable elements. Given the underlying first-order phase transition between the condensed and gas-like states, our experiment may be a realization of self-organized bistability \cite{DiSanto2016}. The inter-state switching is facilitated by both quenched disorder and dynamical noise. The time scales of individual stable and unstable periods are not symmetric (see Fig.\ S3). Durations of instability are mostly determined by the damping time, whereas the stability durations can be much longer and depend on the nucleation of an energy-redistribution event. This is a common property in many excitable systems \cite{Lindner2004}, where the relaxation path is more deterministic than the excitation path. However, the distribution of switching timescales in the experiment is narrower than in the simulation. This may be due to a weak periodic signal in the experiment which couples with the noise to induce switching. The source of the periodicity, in addition to controlling the vertical oscillations through modulating the electrode voltage, are subjects of current investigation in our lab. 

\textit{Acknowledgments} --- This work was supported by the NSF DMR Grant No. 1455086. We thank E. R. Weeks, K. Wiesenfeld, R. Khomeriki, J. Goree, I. Nemenman, G. H. Hentschel, F. Family, D. Hofmann, B. Beal, and C. Carter for fruitful discussions.
 
\bibliography{nsrbibl}

\end{document}